\documentclass[11pt]{article}

\usepackage[letterpaper,portrait,margin=1in]{geometry}
\usepackage{mathpazo}
\usepackage{amsmath,amssymb,bm}
\usepackage{booktabs}
\usepackage{caption}
\usepackage{float}
\usepackage{graphicx}
\usepackage{microtype}
\usepackage{natbib}
\usepackage{paralist}
\usepackage{tcolorbox}
\tcbuselibrary{breakable}
\usepackage{xcolor}
\newcommand{\textred}[1]{\textcolor{red}{\bf #1}}
\usepackage{fancyhdr}
\usepackage{url}
\usepackage[draft]{hyperref}
\hypersetup{hidelinks}

\bibpunct{(}{)}{;}{a}{}{,}
\DeclareGraphicsExtensions{.pdf,.png,.jpg,.eps}

\definecolor{boxcolor}{RGB}{193,229,255}

\newcounter{aff}
\newcommand{\affil}[2]{%
  \par\noindent\refstepcounter{aff}\label{#1}%
  \textsuperscript{\theaff}\textit{#2}\par}
\newcommand{\affmark}[1]{\textsuperscript{\ref{#1}}}

\newcommand{\Msun}{\mathrm{M}_{\odot}}

\newcommand{\Mmol}{M_{\mathrm{mol}}}
\newcommand{\Mdyn}{M_{\mathrm{dyn}}}
\newcommand{\Mstar}{M_{\star}}
\newcommand{\MBH}{M_{\mathrm{BH}}}
\newcommand{\aco}{\alpha_{\mathrm{CO}}}
\newcommand{\acounit}{\,\Msun\,(\mathrm{K\,km\,s^{-1}\,pc^{2}})^{-1}}
\newcommand{\micron}{\mbox{$\mu$m}}

\renewcommand{\textred}[1]{{#1}}

\begin{document}

\noindent
\begin{raggedright}
{\Huge\bfseries A Vision for Little Red Dots in the Local Universe with ALMA2040
\par}\end{raggedright}

\vspace{1.0em}

\noindent
Fengwu Sun\affmark{aff:westlake}$^{,\ast}$, Xiaojing Lin\affmark{aff:tsinghua}, Miguel Pereira-Santaella\affmark{aff:csic}, Dieu D.\ Nguyen\affmark{aff:michigan}

\vspace{1em}
{\footnotesize{
\affil{aff:westlake}{Department of Astronomy, School of Science, Westlake University, Hangzhou, Zhejiang 310030, P.\ R.\ China}
\affil{aff:tsinghua}{Department of Astronomy, Tsinghua University, Beijing 100084, P.\ R.\ China}
\affil{aff:csic}{Instituto de F\'isica Fundamental, CSIC, Calle Serrano 123, 28006 Madrid, Spain}
\affil{aff:michigan}{Department of Astronomy, University of Michigan, 1085 South University Avenue, Ann Arbor, MI 48109, USA}
\noindent $^\ast$\textbf{Corresponding author: \href{mailto:sunfengwu@westlake.edu.cn}{sunfengwu@westlake.edu.cn}}\\
}}

\begin{tcolorbox}[breakable,boxsep=3pt,left=0pt,right=0pt,top=1pt,bottom=1pt,colframe=boxcolor,colback=boxcolor]
\noindent\textbf{Abstract.}
The discovery of Little Red Dots (LRDs) at high redshifts with JWST has revealed a previously unknown phase of black-hole growth. 
However, their black-hole (BH) masses, BH-to-stellar mass ratios, cold-gas reservoirs, and depletion times remain uncertain.
The recent discovery of low-redshift LRDs at $z\simeq0.1 - 0.9$ creates a decisive opportunity to access their low-$J$ CO lines for measuring molecular gas reservoirs, and the nearest systems can in principle be spatially resolved.
We use J1025+1402 (``The Egg'') at $z=0.1007$, the best-studied local LRD, as a benchmark to explore the scientific opportunities with ALMA2040.
Assuming a molecular gas mass of $10^9\,\Msun$, the CO(2--1) detectability strongly depends on the assumed CO-to-H$_2$ conversion ($\aco$).
With a large $\aco \simeq 315 - 725\acounit$ at a low gas-phase metallicity, the CO line would remain very challenging for the current ALMA to detect, while ALMA2040 will likely deliver a robust CO detection and thus a gas-reservoir constraint for The Egg.
Furthermore, with the best angular resolution at its CO(2--1) frequency (14.6\,mas, or 27\,pc), ALMA2040 may spatially resolve the CO emission from The Egg.
Assuming an optimistic, SMC-like $\aco$ with a 50-hour deep integration, we produce mock position--velocity diagrams that may distinguish a $10^7\,\Msun$ black hole from the no-BH model through its steeper central velocity structure.
We conclude that ALMA2040 has the potential to directly measure the gas masses, depletion times, and excitation of local LRDs, while the CO-bright subset may ultimately yield dynamical constraints on the mass of their central black holes.
\end{tcolorbox}

\section{Scientific Context and Open Questions}

\subsection{Little Red Dots: unknowns in black hole mass and cold gas content}

The identification of Little Red Dots (LRDs) at high redshifts (primarily $z \simeq 4 - 8$) marks one of the major discoveries from JWST.
Despite the heterogeneous selection criteria adopted in the literature, LRDs are commonly characterized by their compact red optical continuum emission, strong Balmer breaks, and broad Balmer emission \citep[e.g.,][]{Matthee2024,Greene2024,Kocevski2024,Maiolino2024,Harikane2024}.
The presence of broad lines and compact sizes strongly implicates accreting black holes (BHs), generally at much lower luminosities and inferred masses than those of the luminous quasars previously known at the same epochs.
At the same time, LRDs remain mostly undetected in deep archival X-ray, far-infrared, and radio observations, making them distinct from the vast majority of active galactic nuclei (AGNs) known before JWST.
Dense, dust-poor gas surrounding the accreting black holes has consequently become a leading ingredient in explanations of the Balmer absorption, continuum breaks, and nearly thermal continua that are now frequently observed among LRDs \citep[e.g.,][]{Naidu2025, deGraaff2025, Rusakov2026, InayoshiHo2026}.

There are many fundamental unknowns regarding the nature of LRDs.
Among them, one of the most critical debates concerns the black-hole masses of LRDs.
With single-epoch empirical broad-H$\alpha$ (or H$\beta$) virial estimators, LRDs would host black holes with masses $\MBH \gtrsim 10^7\,\Msun$ that accrete with Eddington ratio $\lambda_\mathrm{Edd} \simeq 0.1 - 1$.
Under such an assumption, LRDs would host overly massive black holes with BH-to-stellar mass ratios approaching $\MBH/\Mstar\sim0.1$ \citep[e.g.,][]{Harikane2024,Maiolino2024}.
This is apparently far above the $\MBH - \Mstar$ relation occupied by the majority of AGNs in the local Universe \citep[e.g.,][]{GreeneHo2005, ReinesVolonteri2015}.
In an alternative interpretation, the wings of broad H$\alpha$ lines are produced partly by electron scattering in a dense ionized envelope.
With such a model, the intrinsic velocity dispersion of broad Balmer lines is much narrower, and the inferred BH masses are typically about two orders of magnitude below the aforementioned ``uncorrected'' virial estimates \citep{Rusakov2026}.
This white paper does not attempt to settle the ongoing debate through line-profile arguments \citep[e.g.,][]{Scholtz2026,Madau2026}.
Instead, we highlight the difficulty in directly measuring the BH mass of LRDs with current facilities \citep[see][]{Juodzbalis2026-nature}.

On the other hand, although evidence of a dense gas envelope has been routinely reported for LRDs \citep[e.g.,][]{Ji-xihan2025, Taylor2025, Kokorev2026}, the cold, molecular gas content of the host galaxies of LRDs is almost entirely unknown.
So far, the vast majority of LRDs are not detected with ALMA or NOEMA observations either in dust continuum \citep[e.g.,][]{Setton2025, Casey2025} or in emission lines as gas tracers \citep[e.g.,][]{Xiao2025,Akins2026}.
To our knowledge, the only exception is A383-LRD1 at $z=6.027$ \citep{Golubchik2026, 2026ApJ..1006L..14B}, for which ALMA [C~II] 158\,\micron\ and [O~III]\,88\,\micron\ detections have been reported \citep{Knudsen2016, 2025arXiv251205097K}.
However, we note that the A383-LRD1 system consists of a red LRD and a UV-bright blue companion separated by only $\sim0.3''$, making the physical origin of the lines ambiguous.
Without the direct detection of any molecular gas tracer, none of the molecular gas mass $\Mmol$, the gas depletion time $t_{\rm dep}=\Mmol/{\rm SFR}$, or the physically relevant ratio $\MBH/(\Mstar+\Mmol)$ can be determined.
These quantities are key to diagnosing the interstellar medium (ISM) of the LRD host galaxies, which sets the gas fuel supply for the central BH growth.
Measurements of several [C~II], [C~I], or CO transitions would additionally test the temperature and density of the gas in LRD galaxies, and may eventually provide a direct observational link to the coevolution (or not) of the BH and host galaxy of an LRD.

\subsection{The solutions emerging at low redshifts}

\begin{figure}[!t]
\centering
\includegraphics[width=\linewidth]{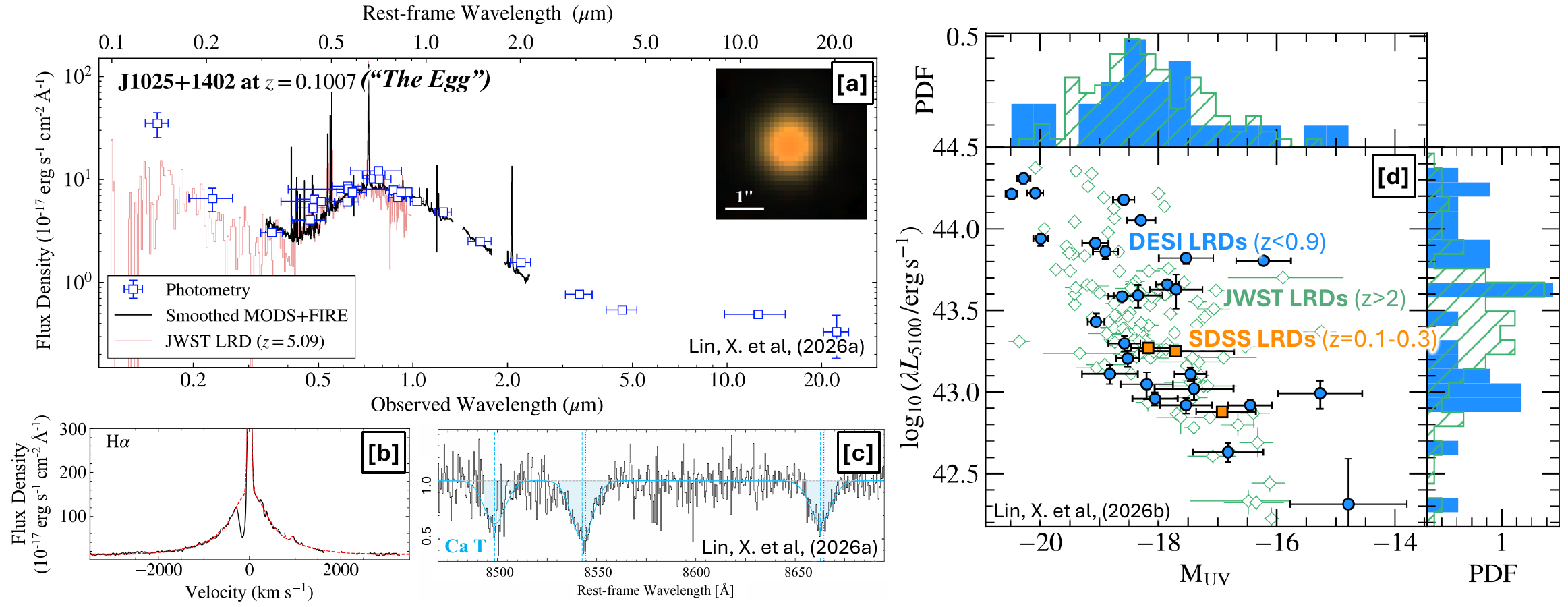}
\caption{
\textsl{[a]}: The SED and optical spectrum of J1025+1402 (``The Egg'') at $z=0.1007$.
For comparison, a JWST NIRSpec spectrum of an LRD at $z=5.09$ is also shown.
The SDSS image cutout is also shown.
\textsl{[b]}: Zoom-in view of the H$\alpha$ broad lines. Absorption features are clearly seen, similar to those of high-redshift LRDs.
\textsl{[c]}: Zoom-in view of the Ca\,II triplet absorption lines. The extreme Ca\,T absorption exceeds the strengths of all known AGNs, galaxies and most stars, strongly supporting the presence of a dense gas envelope.
\textsl{[d]}: Rest-frame optical (5100\,\AA) versus UV luminosity of LRDs selected at high redshifts with JWST (green; \citealt{deGraaff2025b}) and at low redshifts from SDSS (orange) and DESI DR1 (blue), occupying the same parameter space.
Figure adapted from \citet{Lin2026, Lin2026b-arXiv260521574L}.
}
\label{fig:connection}
\end{figure}

A key recent breakthrough in the study of LRDs is the identification of LRDs in the low-redshift Universe ($z \lesssim 0.2$).
\citet{Lin2026} presented three LRDs at $z\simeq0.1 - 0.2$ identified among roughly three million SDSS spectra (Figure~\ref{fig:connection}).
These low-redshift LRDs share all the defining features of their high-redshift counterparts, including a point-like morphology, a V-shaped UV-to-near-IR spectral energy distribution (SED), broad Balmer emission, Balmer absorption superposed on the broad lines, weak X-ray emission, and a lack of optical variability.
These features, especially the presence of Balmer absorption lines and unusually strong Ca~II triplet absorption, provide direct evidence for a cool, dense gas envelope.
These absorption features have been shown to be extremely rare among low-redshift AGNs \citep[e.g.,][]{Shangguan2026}.
These strong absorptions undoubtedly make these objects distinct from other types of sources that may also show V-shaped SEDs at UV-to-optical wavelengths, for example, dwarf galaxies or rejuvenated quiescent galaxies containing AGNs.
The nearest and best-studied source in this sample, J1025+1402 at $z=0.1007$, is nicknamed ``The Egg'' and was independently analyzed as the ``Lord of LRDs'' by \citet{Ji2026}.
\textred{\citet{Lin2026} inferred $\log(\MBH/\Msun)=6.25^{+0.08}_{-0.07}$ for The Egg, while \citet{Ji2026} found a systematic range of $5.95$--$7.22$ assuming a variety of scenarios (e.g., virial relations and electron scattering).
The stellar mass inferred from SED fitting, $\Mstar=10^{9.9\pm0.1}\,\Msun$, likely includes substantial AGN contamination and exceeds the dynamical limit $\Mdyn<3.3\times10^9\,\Msun$ \citep{Ji2026}. 
As a consequence, we adopt an illustrative $\Mstar=10^9\,\Msun$ below.}

The known population of low-redshift LRDs is now undergoing substantial growth.
\citet{Park2026} identified an additional eight LRDs at $z = 0.2 - 0.45$ in DESI DR1, and an expanded search by \citet{Lin2026b-arXiv260521574L} \textred{increases} the DESI LRD sample at $z = 0.2 - 0.9$ to $N = 27$.
Ignoring redshift, the low- and high-redshift samples of LRDs occupy strikingly similar regions in broad-line luminosity, continuum luminosity, Balmer-break strength, Balmer decrement, line vs.\ continuum luminosity relation, narrow-line diagnostics (e.g., BPT diagrams), and gas-phase metallicity \citep{Lin2026b-arXiv260521574L}.
The key advantage of the low-redshift LRD sample is their brightness: being $\gtrsim100\times$ brighter than their high-redshift counterparts, these low-redshift LRDs are ideal laboratories for characterizing the host galaxy properties and even exploring the BH masses through direct, kinematic measurements.

\subsection{Open questions that need to be addressed by ALMA2040}

Existing ALMA observations detect nearly no continuum or emission lines for the vast majority of LRDs at $z \gtrsim 4$ \citep[e.g.,][]{Casey2025,Setton2025,Xiao2025,Akins2026}, clearly demonstrating the necessity of the local route.
At $z\gtrsim4$, the low-$J$ CO transitions (e.g., CO(3--2) and CO(2--1)) that most directly trace bulk molecular gas are redshifted below ALMA's frequency coverage, or remain very challenging to observe with ALMA Band-2/1.
[C~II]\,158\,\micron\ is accessible, but the strength of the [C~II] emission depends on a variety of physical conditions within the ISM of the LRD host galaxies, including ionization state, radiation field, and metallicity.
An upper limit on [C~II] luminosity is therefore not equivalent to a clean molecular-mass limit.
In contrast, LRDs at low redshifts benefit from (\romannumeral1) a $\gtrsim 100\times$ boost in brightness;
(\romannumeral2) $\sim2\times$ better physical resolution, comparing the angular diameter distances at $z=0.2$ and $z = 5$;
(\romannumeral3) access to low-$J$ CO emission in the most sensitive ALMA bands (e.g., Band-6/7).
Observations with ALMA2040 will address the following key open questions regarding the cold gas content and BH--host coevolution of LRDs:
\begin{compactitem}
\item Do LRDs contain reasonably sized molecular reservoirs (e.g., $\Mmol\sim10^9\,\Msun$) expected for sustained black-hole growth and star-formation activity, or are they already poor in cold molecular gas?
\item Are their gas depletion times and CO excitation consistent with those measured for compact star-forming galaxies, or galaxies undergoing obvious AGN heating and/or feedback?
\item By taking into account the gas masses or directly obtaining dynamical masses through resolved ALMA spectral cubes, do LRDs have $\MBH/(\Mstar+\Mmol)$ or $\MBH/\Mdyn$ ratios consistent with those measured for nearby galaxies?
\item Can resolved molecular-gas dynamics independently distinguish a $\sim10^7\,\Msun$ black hole from the lower masses (e.g., $\sim10^6\,\Msun$) implied by electron-scattering interpretations?
\end{compactitem}

\section{ALMA2040 Technical Requirements}

\subsection{A ten-hour CO(2--1) experiment on The Egg}

We test the capability of ALMA2040 in studying LRDs at low redshifts with ``The Egg'' as a benchmark.
We assume a flat $\Lambda$CDM cosmology with $H_0=70$ km s$^{-1}$ Mpc$^{-1}$ and $\Omega_{\rm m}=0.3$.
At $z=0.1007$, the CO(2--1) emission from ``The Egg'' is observed at 209.45 GHz and the luminosity distance is 463.7 Mpc.
For the sensitivity calculation, we assume a total molecular mass $\Mmol=10^9\,\Msun$, a 100~km\,s$^{-1}$ line width, and an unresolved source size for the detection experiment.
\textred{This benchmark gives a molecular gas fraction $f_{\rm mol}=\Mmol/(\Mstar+\Mmol)$ of 0.5 and a total baryonic mass below the conservative dynamical mass upper limit.}

In this calculation, the dominant astrophysical uncertainty is the CO-to-H$_2$ conversion factor.
A Milky-Way value, $\aco=4.35\acounit$ with $r_{21}=0.6$, predicts a bright $S_{21}\Delta v=1.15$\,Jy\,km\,s$^{-1}$, which would be very easy for the current ALMA to detect.
This is likely too optimistic because the gas-phase metallicity of The Egg is only $12+\log({\rm O/H})=7.55\pm0.03$ (i.e., $7\%$ solar metallicity), inferred from its narrow-line emission; \citealt{Lin2026}, and $\alpha_\mathrm{CO}$ will sharply increase in low-metallicity environments \citep[e.g., $\alpha_\mathrm{CO} \approx 70\acounit$ for SMC at $12+\log({\rm O/H})\approx 8.0$;][]{2011ApJ...737...12L}.
We therefore also perform the sensitivity calculation using the metallicity-dependent prescription of \citet{Accurso2017},
\begin{equation}
\log\aco=14.752-1.623[12+\log({\rm O/H})]+0.062\log\Delta({\rm MS}).
\end{equation}
Because the stellar mass and the star-forming main-sequence offset of The Egg are poorly constrained, we neglect the weak secondary dependence on $\Delta({\rm MS})$ by setting $\log\Delta({\rm MS})=0$.
We note that the \citet{Accurso2017} relation is calibrated only at $12+\log({\rm O/H})>7.9$, and therefore extrapolating it to the measured metallicity of The Egg gives $\aco\simeq315\acounit$.
As an empirical benchmark, the dust-based Local Group measurements of \citet{2011ApJ...737...12L} give $\aco\simeq70\acounit$ for the SMC at $12+\log({\rm O/H})\simeq8.0$.
For comparison, the shielding-based prescription summarized by \citet{Bolatto2013}, $\aco\simeq2.9\exp[0.4/(Z'\Sigma_{\rm GMC}^{100})]$, gives $\aco\simeq725\acounit$ at the metallicity of The Egg and the reference cloud surface density $\Sigma_{\rm GMC}^{100}=1$.
We therefore include the SMC value as an empirical intermediate benchmark and retain $\aco=315$--$725\acounit$ as an illustrative bracket extrapolated to the lower metallicity of The Egg.
Adopting $r_{21}=0.6$, these assumptions give a CO(2--1) line flux of 0.0717~Jy\,km\,s$^{-1}$ for the SMC benchmark and 0.0069--0.0159~Jy\,km\,s$^{-1}$ for the Egg-metallicity extrapolations.

Based on the expected ALMA2040 sensitivity table, we estimate an RMS noise of 54.5~$\mu$Jy\,beam$^{-1}$ through one-hour on-source integration per 10\,km\,s$^{-1}$ channel at 209.4\,GHz.
With ten hours of integration over a 100~km\,s$^{-1}$ line width, we derive an RMS noise for the integrated line flux of $5.45\times10^{-4}$~Jy\,km\,s$^{-1}$.
Table~\ref{tab:detectability} compares this emission line sensitivity with that of current ALMA and presents the expected line signal-to-noise ratio (S/N).

\begin{table}[t]
\centering
\caption{CO(2--1) detectability for The Egg assuming $\Mmol=10^9\,\Msun$, a line width of 100~km\,s$^{-1}$, and a gas-phase metallicity of $12+\log({\rm O/H})=7.55$. $\aco$ is in units of $\Msun\,(\mathrm{K\,km\,s^{-1}\,pc^{2}})^{-1}$.}
\label{tab:detectability}
\small
\begin{tabular}{lcccc}
\toprule
Conversion assumption & $\aco$ & $S\Delta v$ & Current ALMA & ALMA2040 \\
 & & (Jy km s$^{-1}$) & 10 hr S/N & 10 hr S/N \\
\midrule
Milky Way & 4.35 & 1.155 & 227 & 2119 \\
 SMC & 70 & 0.0717 & 14.1 & 131.7 \\
 \citet{Accurso2017} & 315 & 0.0159 & 3.1 & 29.3 \\
 \citet{Bolatto2013} & 725 & 0.0069 & 1.4 & 12.7 \\
\bottomrule
\end{tabular}
\end{table}

Our calculation suggests that a ten-hour ALMA2040 observation would detect the SMC benchmark at approximately $132\sigma$ and the two Egg-metallicity extrapolations at $13$--$29\sigma$ (Figure~\ref{fig:requirements}, left).
The corresponding $5\sigma$ molecular-mass sensitivity is approximately $(1.7$--$3.9)\times10^8\,\Msun$, conditional on the adopted CO-to-H$_2$ conversion.
Current ALMA would detect the SMC benchmark at approximately $14\sigma$, but would reach only $1.4$--$3.1\sigma$ for the two Egg-metallicity extrapolations in ten hours and would struggle to provide useful constraints on either the gas mass or the velocity profile.
\textred{For $\Mmol=10^9\,\Msun$, a $5\sigma$ detection with current ALMA in ten hours would require $\aco\lesssim200\acounit$.}

Assuming that the UV emission of The Egg predominantly arises from its host galaxy, we estimate an ${\rm SFR}\simeq0.4\,\Msun\,{\rm yr}^{-1}$.
The benchmark gas mass $10^9\,\Msun$ then corresponds to a depletion timescale of approximately 2.5\,Gyr, while the ALMA2040 detection limit corresponds to a depletion time limit of $\sim0.4 - 1.0$ Gyr.
\textred{In principle, a dust continuum measurement could validate the CO-based gas mass, although the uncertain gas-to-dust ratio at $\simeq7\%$ solar metallicity and possible AGN, free--free, or synchrotron contamination make it very difficult to provide a unique gas-mass estimate \citep{2014A&A...563A..31R,2016ApJ...820...83S}.}

Taking The Egg as an example, we conclude that ALMA2040 will likely deliver the robust low-$J$ CO line detections and emission-line profile measurements that remain inaccessible or challenging for ALMA today.
Once any CO transition is detected, the remaining CO transitions (and [C~I] emission) will become natural follow-up targets to measure the first LRD CO spectral-line energy distributions (SLEDs).
The detectability of other lines is a function of the gas excitation and the frequency-dependent sensitivity, and the CO SLED will be the key to distinguishing a cold gas reservoir from highly excited gas \citep{CarilliWalter2013} and assessing the role of the accreting black hole in providing feedback on the surrounding ISM.

\subsection{From detection to a dynamical black-hole test}

ALMA2040 may enable a second, more challenging experiment: resolved gas dynamics.
At The Egg's redshift, one arcsecond corresponds to 1.86\,kpc, and therefore an exponential gas disk with a circularized effective radius of $r_{\rm e}=100$\,pc subtends 54\,mas.
The ALMA2040 extended reference configuration will offer a 14.6\,mas synthesized beam size at 209 GHz (or 27 pc), placing approximately 3.7 beams across $r_{\rm e}$.
For an illustrative model, we adopt $\Mmol=10^9\,\Msun$ for the exponential gas disk and a separate $\Mstar=10^9\,\Msun$ stellar component with $R_{\rm e,star}=500$ pc under a Hernquist profile.
We assume an inclination $i=60^\circ$, and intrinsic gas velocity dispersion $\sigma_{\rm gas}=20$ km s$^{-1}$.
We note that the stellar size is motivated by the HST-resolved host galaxy size of J1047+0739 ($R_{\rm e}=530$ pc; \citealt{Lin2026}) and is not a direct measurement of The Egg.

For a characteristic stellar dispersion of 50 km s$^{-1}$, the black-hole sphere of influence is only 1.7 pc for $\MBH=10^6\,\Msun$ and 17 pc for $\MBH=10^7\,\Msun$.
Neither is formally resolved by a 27 pc beam, but the larger black hole changes the projected circular speed at half a beam from about 56 to 74 km s$^{-1}$ in this model.
This $\sim18$ km s$^{-1}$ contrast is a geometric prediction and does not by itself establish detectability after surface-brightness dilution and beam smearing.
At 14.6 mas resolution, the adopted exponential disk spans roughly 100 effective synthesized beams, reducing the fiducial $13 - 29\sigma$ unresolved detection to only $1 - 3\sigma$ in a spatially matched, line-integrated high-resolution measurement.
The estimated central-beam S/N in an individual 10 km s$^{-1}$ channel is below unity.
Therefore, under the fiducial low-metallicity conversion factors, a ten-hour observation will not provide a robust black-hole dynamical measurement.

\begin{figure}[!t]
\centering
\includegraphics[width=0.49\linewidth]{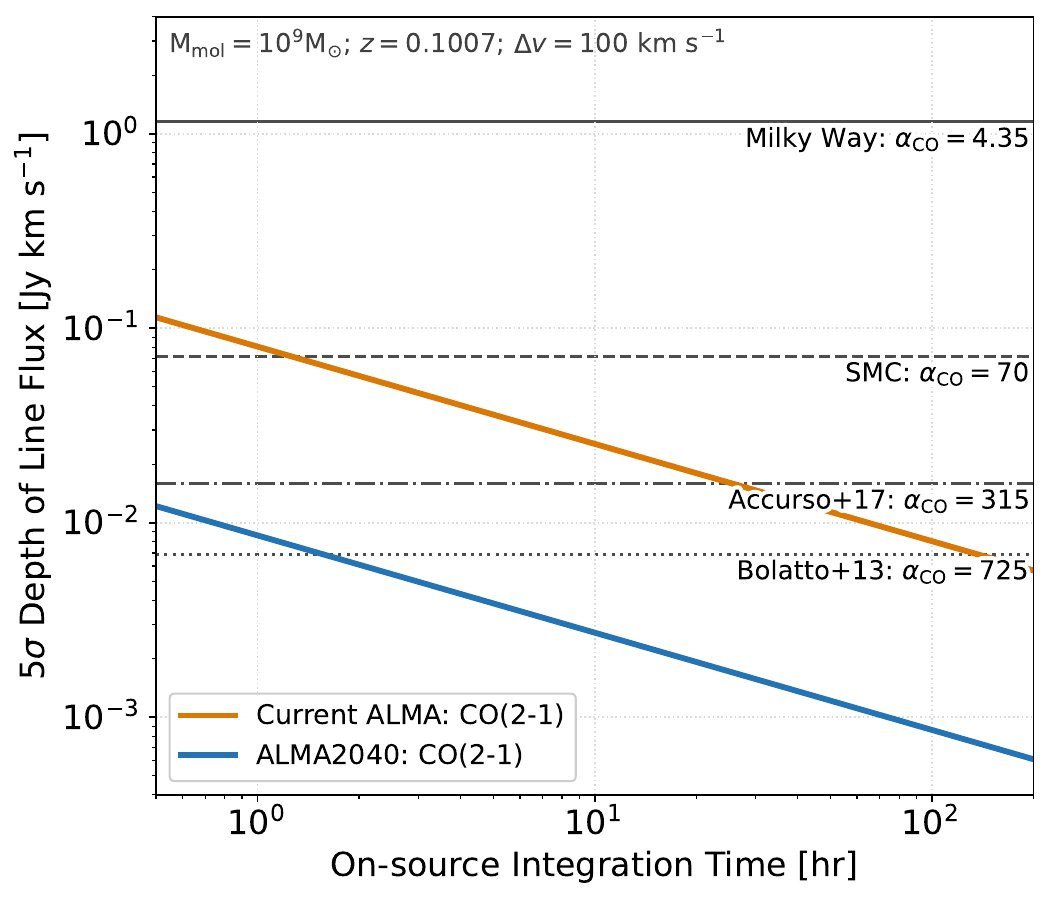}
\includegraphics[width=0.49\linewidth]{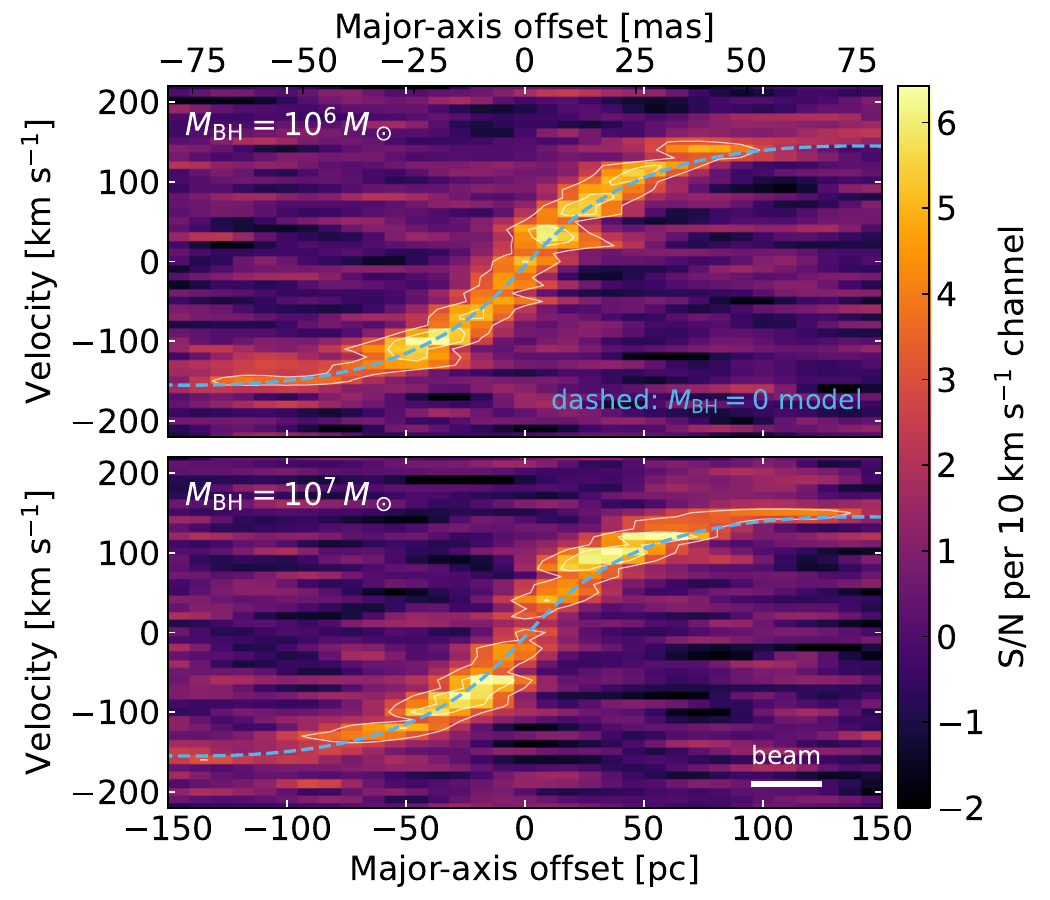}
\caption{\textsl{Left}: $5\sigma$ CO(2--1) line detection limit at the redshift of The Egg with current ALMA (orange) and ALMA2040 (blue), assuming a line width of 100~km\,s$^{-1}$.
For a molecular gas mass of $\Mmol = 10^9\,\Msun$, the expected line fluxes for various $\aco$ are shown as horizontal lines.
\textsl{Right}: Simulated position--velocity diagrams of The Egg as observed with ALMA2040, assuming $\aco = 70\acounit$ and a 50-hour on-source integration (see the main text for the other assumptions of the gas and stellar profiles).
Contours denote the $3\sigma$ and $5\sigma$ levels per 10~km\,s$^{-1}$ channel.
Under such idealized assumptions, the $\MBH = 10^6\,\Msun$ scenario cannot be differentiated from the null ($\MBH = 0$) model, while the $\MBH = 10^7\,\Msun$ scenario will exhibit high-velocity wings at $r < 50$\,pc, hinting at the presence of an overly massive BH ($\MBH/\Mstar = 0.01$).
}
\label{fig:requirements}
\end{figure}

However, if the CO emission is substantially brighter than our fiducial estimate, for example at the SMC-like $\aco \simeq 70\acounit$ benchmark, a deeper 50-hour integration would raise the spatially matched high-resolution significance to $\sim28\sigma$, with an estimated central 10 km s$^{-1}$ channel significance of $\sim7\sigma$.
We simulate mock position--velocity diagrams under these assumptions in Figure~\ref{fig:requirements} (right), where the $\MBH = 10^7\,\Msun$ case is distinguishable from the null model through the steeper central velocity structure.
By forward modeling the full spectral cube, we may then constrain the dynamical mass and test for the presence of a $M_\mathrm{BH} \sim10^7\,\Msun$ black hole, although the result will remain sensitive to beam smearing, inclination, and the stellar and gas mass profiles.
Such an optimistic scenario could arise if the extrapolated $\aco$ is indeed too large, if the CO is substantially more compact than our assumption, or if another low-redshift LRD is intrinsically more suitable for resolved ALMA follow-up than The Egg.
Again, an $M_\mathrm{BH} \sim 10^6\,\Msun$ black hole will remain inaccessible at this spatial resolution.

A robust dynamical-mass measurement of LRDs therefore requires selecting high-surface-brightness targets before obtaining deep ALMA2040 integration with the longest available baselines.
A compact or intermediate-resolution array component will need to be included to recover the total flux and constrain the extended mass distribution.
As a result, observations of the low-redshift LRD population with the current ALMA are essential for the preparation of ALMA2040.
In ALMA Cycle 13, programs such as {2026.1.00137.S} (PI: D'Eugenio, F.) and {2026.1.00880.S} (PI: Kiyota, T.) have been accepted to observe two low-redshift LRDs, J1025+1402 and J1047+0739, providing the first pathfinder constraints.
\textred{Program 2026.1.00137.S will observe J1047+0739 for Band~2 continuum, while 2026.1.00880.S will observe the CO(3--2) emission of The Egg down to a nominal gas-mass sensitivity of $\sim10^8\,\Msun$ under relatively optimistic CO-excitation and $\alpha_\mathrm{CO}$ assumptions.}
\textred{Neither program will target CO(2--1), but both programs will inform ALMA2040 target selection.}
We also highlight that observations of a much larger low-redshift LRD sample in future ALMA cycles will be essential to identify CO-bright sources and determine the best targets for resolved ALMA2040 follow-up.

\section{Synergies with current and future facilities}

Currently, the sample size of known low-redshift LRDs remains small, as the typical rate of LRD identification is $\sim$1 per million SDSS or DESI spectra \citep{Lin2026,Lin2026b-arXiv260521574L}.
Enlarging the sample size would be one of the most direct preparations for ALMA2040.
Because of the extreme rarity, wide-field spectroscopic surveys are the most efficient and scalable discovery route for low-redshift LRDs.
The ongoing Stage-IV cosmology survey DESI \citep{DESI2016} has already delivered the first sizable sample \citep{Park2026,Lin2026b-arXiv260521574L}, and ongoing or upcoming multi-object spectroscopic facilities such as Subaru/PFS \citep{Takada2014}, VISTA/4MOST \citep{deJong2019}, and VLT/MOONS \citep{Cirasuolo2020} will extend the search, especially to the southern sky.
Wide-field space missions, including Euclid \citep{Euclid2025}, Roman \citep{ROTAC2025}, CSST \citep{CSST2026}, and SPHEREx \citep{Bock2026}, will provide slitless spectroscopy and all-sky spectrophotometry that are sensitive to the distinctive V-shaped SEDs and strong Balmer line features of LRDs.
Looking further ahead, the proposed Stage-V spectroscopic facilities of the 2030s, such as Spec-S5 \citep{Besuner2025} and MUST \citep{Zhao2024}, anticipate order-of-magnitude larger spectroscopic samples.
Together, these surveys will expand the low-redshift LRD census from tens to potentially hundreds of sources, from which the rare CO-bright and dynamically favorable targets for ALMA2040 may be identified.

In parallel, high-spatial-resolution imaging and spectroscopy from the rest-frame UV to radio wavelengths will sharpen our understanding of the LRD host galaxies.
Deep HST and JWST observations have been scheduled for a small subset of low-redshift LRDs to measure the structures and stellar masses of their host galaxies, which will anchor the $\MBH/\Mstar$ and $\MBH/(\Mstar+\Mmol)$ comparisons enabled by ALMA2040.
Near-infrared interferometry with VLT/GRAVITY+ \citep{GravityPlus2022} may spatially resolve the broad-line kinematics of the brightest low-redshift LRDs through Paschen lines, as demonstrated for luminous quasars \citep{Gravity2018}, once a suitable adaptive-optics reference star is available.
Diffraction-limited imaging with ELT/MICADO \citep{Davies2021} will resolve the stellar hosts at $\sim10$ mas resolution, comparable to the extended-configuration ALMA2040 beam.
At radio wavelengths, the ngVLA \citep{Murphy2018} will constrain any compact jet or free--free emission and provide an extinction-free view of the nuclear activity.
Dynamical BH masses inferred from ALMA2040 CO kinematics, GRAVITY+ broad-line interferometry, and stellar or ionized-gas kinematics from ELT spectroscopy would then serve as independent cross-checks on the same objects, turning the current BH-mass debate into a solvable measurement problem.

\section*{Acknowledgements}
We thank Roberto Decarli for organizing the efforts of the ALMA2040 AGN Working Group.
We thank Jacqueline Hodge and Umberto Maio for their helpful comments.
We thank Roberto Gilli for his careful and helpful review.
This work was developed as part of the broader ALMA2040 community effort.
We acknowledge the Lorentz Center and NOVA for supporting the community workshops and coordination activities that helped initiate and shape the ALMA2040 White Paper process.
The authors acknowledge the use of generative artificial intelligence (GPT 5.6 Sol and Claude Fable 5) in preparing this manuscript, under human supervision and review.
The authors are solely responsible for all content presented in this manuscript.

\bibliographystyle{apj_mod}
\bibliography{main}

\end{document}